\documentclass[singlecolumn,secnumarabic,amssymb, aps, preprint,superscriptaddress]{revtex4-2}
\usepackage{float}
\usepackage{bigints}
\usepackage[parfill]{parskip}  
\usepackage{dirtytalk}
\usepackage{amsmath}
\usepackage{caption}
\usepackage{amssymb}
\usepackage{mathrsfs}
\usepackage{amsthm}
\usepackage{braket}
\usepackage{esint}
\usepackage{setspace}
\usepackage{multirow}
\usepackage{rotating}
\usepackage{comment}
\usepackage{dcolumn}
\usepackage{bm}%
\usepackage[dvipsnames]{xcolor}
\setcitestyle{super,open={},close={}}

\begin{document}

\title{Spontaneous Polarization Induced Photovoltaic Effect In Rhombohedrally Stacked $\text{MoS}_2$}

\author{$\text{Dongyang Yang}^*$}
 \affiliation{Department of Physics and Astronomy, The University of British Columbia,Vancouver,BC V6T 1Z1, Canada}
 \affiliation{Quantum Matter Institute, The University of British Columbia,Vancouver,BC V6T 1Z4, Canada}
\author{$\text{Jingda Wu}^*$}
 \affiliation{Department of Physics and Astronomy, The University of British Columbia,Vancouver,BC V6T 1Z1, Canada}
 \affiliation{Quantum Matter Institute, The University of British Columbia,Vancouver,BC V6T 1Z4, Canada}
 
 \author{Benjamin T. Zhou}
 \affiliation{Department of Physics and Astronomy, The University of British Columbia,Vancouver,BC V6T 1Z1, Canada}
 \affiliation{Quantum Matter Institute, The University of British Columbia,Vancouver,BC V6T 1Z4, Canada}
 
 \author{Jing Liang}
 \affiliation{Department of Physics and Astronomy, The University of British Columbia,Vancouver,BC V6T 1Z1, Canada}
 \affiliation{Quantum Matter Institute, The University of British Columbia,Vancouver,BC V6T 1Z4, Canada}
 
\author{Toshiya Ideue}
 \affiliation{Quantum-Phase Electronics Center (QPEC) and Department of Applied Physics, The University of Tokyo, Tokyo 113-8656, Japan}
 \affiliation{RIKEN Center for Emergent Matter Science (CEMS), Wako, Saitama 351-0198, Japan}
 \author{Teri Siu}
 \affiliation{Quantum Matter Institute, The University of British Columbia,Vancouver,BC V6T 1Z4, Canada}
 \author{Kashif Masud Awan}
 \affiliation{Quantum Matter Institute, The University of British Columbia,Vancouver,BC V6T 1Z4, Canada}
 \author{Kenji Watanabe}
 \affiliation{Research Center for Functional Materials,
National Institute for Materials Science, 1-1 Namiki, Tsukuba 305-0044, Japan}
 \author{Takashi Taniguchi}
 \affiliation{International Center for Materials Nanoarchitectonics,
National Institute for Materials Science,  1-1 Namiki, Tsukuba 305-0044, Japan}
 \author{Yoshihiro Iwasa}
 \affiliation{Quantum-Phase Electronics Center (QPEC) and Department of Applied Physics, The University of Tokyo, Tokyo 113-8656, Japan}
 \affiliation{RIKEN Center for Emergent Matter Science (CEMS), Wako, Saitama 351-0198, Japan}
\author{Marcel Franz}
 \affiliation{Department of Physics and Astronomy, The University of British Columbia,Vancouver,BC V6T 1Z1, Canada}
 \affiliation{Quantum Matter Institute, The University of British Columbia,Vancouver,BC V6T 1Z4, Canada}
\author{$\text{Ziliang Ye}^{\dag}$}
 \affiliation{Department of Physics and Astronomy, The University of British Columbia,Vancouver,BC V6T 1Z1, Canada}
 \affiliation{Quantum Matter Institute, The University of British Columbia,Vancouver,BC V6T 1Z4, Canada}
 \email{zlye@phas.ubc.ca}

\maketitle

\section*{A\lowercase{bstract}}

Stacking order in van der Waals (vdW) materials determines the coupling between atomic layers and is therefore key to the materials’ properties. By exploring different stacking orders, many novel physical phenomena have been realized in artificial vdW stacks\cite{cao2018unconventional,sharpe2019emergent,tang2020simulation,regan2020mott}. Recently, ferroelectricity, a phenomenon exhibiting reversible spontaneous electrical polarization, has been observed in zero-degree aligned hexagonal boron nitride (hBN)\cite{woods2021charge,yasuda2021stacking,stern2021interfacial} and graphene-hBN heterostructures\cite{zheng2020unconventional}, holding promise in a range of electronic applications\cite{tsymbal2021two}. In those artificial stacks, however, the single domain size is limited by the stacking-angle misalignment to about 0.1 to 1 $\mu$m, which is incompatible with most optical or optoelectronic applications. Here we show molybdenum disulfide ($\text{MoS}_2$) in the rhombohedral phase can host a homogeneous spontaneous polarization throughout few-$\mu$m-sized exfoliated flakes, as it is a natural crystal requiring no stacking and is therefore free of misalignment. Utilizing this homogeneous polarization and its induced depolarization field (DEP), we build a graphene - $\text{MoS}_2$ based photovoltaic (PV) device with high efficiency. The few-layer $\text{MoS}_2$ is thinner than most oxide-based ferroelectric films, which allows us to maximize the DEP and study its impact at the atomically thin limit, while the highly uniform polarization achievable in the commensurate crystal enables a tangible path for up-scaling. The external quantum efficiency (EQE) of our device is up to 16\% at room temperature, over one order larger than the highest efficiency observed in bulk photovoltaic devices\cite{zhang2019enhanced,akamatsu2021van}, owing to the reduced screening in graphene, the exciton-enhanced light-matter interaction, and the ultrafast interlayer relaxation in $\text{MoS}_2$. In view of the wide range of bandgap energy in other transition metal dichalcogenides (TMDs)\cite{mak2016photonics,wang2018colloquium}, our findings make rhombohedral TMDs a promising and versatile candidate for applications such as energy-efficient photo-detection with high speed and programmable polarity.

\section*{M\lowercase{ain}}

Three intrinsic photovoltaic mechanisms exist in polar materials \cite{lopez2016physical}. Two of these mechanisms are second-order effects of the optical field, arising from the shift current and injection current\cite{sipe2000second}. They have recently attracted much attention because of the potential for overcoming the Shockley–Queisser limit in energy harvesting applications\cite{yang2018flexo, zhang2019enhanced,osterhoudt2019colossal,akamatsu2021van}. The third mechanism is related to the depolarization field (DEP). The DEP in a polar material is an electric field generated by the bond charge at the surface or interface where the polarization is terminated. If a polar thin film is sandwiched between two metal electrodes, the DEP is nonzero when the induced image charges in the electrode do not fully compensate the polarization charges\cite{mehta1973depolarization}. Upon excitation, the photocarriers drift under the DEP, forming a PV current. In contrast to the first two, the third mechanism is linearly dependent on the depolarization field and has been observed in oxide-based ferroelectric films\cite{qin2008high,yang2010above,nechache2015bandgap}. Since the DEP's strength depends on the film thickness and screening in  electrodes\cite{qin2008high,batra1973phase,wurfel1973depolarization}, much effort has been dedicated to growing ultra-thin ferroelectric oxide film\cite{fong2004ferroelectricity} and optimizing the contact material\cite{qin2009photovoltaic}. Here we push the DEP effect to the atomically thin limit with few-layer rhombohedral $\text{MoS}_2$ and monolayer graphene as the electrode, which we find can preserve as much as 95\% of the DEP due to its reduced screening in two dimensions.

The rhombohedral stacking order in carbon systems has led to some exotic transport phenomena\cite{chen2019evidence,shi2020electronic,zhou2021superconductivity}. In $\text{MoS}_2$, as the two sublattice sites are not equivalent, the rhombohedral (R) stacking means the neighbouring layers are oriented in the same direction, in contrast with the hexagonal (H) stacking (Fig. 1a). In R-stacking, adjacent layers shift laterally by a third of the unit cell and three shifts complete a unit cell, for which the phase is named 3R\cite{suzuki2014valley,zhao2016atomically}. The relative lateral shift modifies the effective angular momentum at $\pm$K valley  and leads to an asymmetric interlayer coupling between two monolayer $\text{MoS}_2$. As a result, an emergent spontaneous polarization arises along the out of plane direction according to our Berry phase calculation based on an realistic tight-binding model (Details in Supplementary Information).

A schematic band structure of a bilayer 3R-$\text{MoS}_2$ is shown in Fig. 1b. Protected by three-fold rotational symmetry, the Bloch state at the K point in each layer remains almost localized\cite{wang2017interlayer,andersen2021excitons} due to weak interlayer coupling. The interlayer potential induced by the spontaneous polarization and the asymmetric interlayer coupling (Supplementary Information) cause an energy shift in both the conduction and valence bands, forming a type-II band alignment at K point. Since the band offsets are slightly different, the inter-band transitions in the top and bottom layers can be distinguished in energy (Fig. 1c). The extracted A exciton splitting is about 13 meV, in agreement with the measured results in artificial stacks\cite{sung2020broken,andersen2021excitons} and theoretical calculations\cite{wang2017interlayer,kormanyos2018tunable}. One important difference of the bilayer $\text{MoS}_2$ from the monolayer is that its bandgap is indirect\cite{mak2010atomically}, which has a large impact on the PV effect as discussed later.

The schematic of our device structure and band alignment is shown in Fig. 2a. A piece of atomically thin 3R-$\text{MoS}_2$ is sandwiched between two graphene electrodes. Similar structures have been used to build 2H-$\text{MoS}_2$ based photo-detectors\cite{yu2013highly,britnell2013strong}. Unlike the 2H stacking, the spontaneous polarization from the 3R stacking induces the same amount of charge with opposite signs in top and bottom graphene electrodes, thus modifying graphene's chemical potential. (The graphene electrodes are cut from the same original piece to keep initial doping levels the same.) At zero bias, a DEP is established in proportion to the induced potential difference. Due to the small density of state near the Dirac point, the DEP in a graphene-contacted device can be 95\% of that in a completely unscreened film (Supplementary Information). Since the $\text{MoS}_2$ layer is atomically thin (0.7 nm for a monolayer), the Schottky junction effect in our device is negligible, as confirmed by the symmetric current-voltage characteristic (I-V) curve. The DEP drives photocarriers into a photovoltaic current which is measured by a sourcemeter. A graphite back gate is added outside the hBN encapsulation to tune the doping level. All photocurrent measurements are carried out at room temperature.

Our first device (D1) contains a series of terraces including one, two, and three layers of 3R-$\text{MoS}_2$ (Fig. 2b), which is directly exfoliated from a 3R bulk crystal grown by chemical vapor transport method and characterized by both linear and nonlinear optical probes (Methods). First, we obtain a spatial map of the short-circuit current by scanning the sample under laser focus ($\lambda$ = 532 nm) (Fig. 2c). On average, we observe a responsivity of 1.0 $\text{mAW}^{-1}$ in the bilayer, while the photoresponse in the monolayer is about an order of magnitude weaker and close to the noise floor. Moreover, we find an even stronger responsivity with the same polarity in the trilayer (1.5 mA$\text{W}^{-1}$). The photocurrent is independent of the laser polarization, indicating a likely different origin from the shift current or injection current. The sharp contrast between the monolayer and few layers $\text{MoS}_2$ at zero bias suggests an intrinsic PV effect in the rhombohedral device.

To further examine the nature of the PV effect, we study the I-V and the photoresponse spectrum. Under dark condition, we observe a linear I-V curve that crosses the origin, absent of any Schottky behavior from -0.1 mV to 0.1 mV (Fig. 2d). When the laser is focused on the bilayer area, the I-V curve is shifted upward, indicating a negligible photoconductivity effect. As the device resistance is almost unchanged, the phenomenological open-circuit voltage ($V_{oc}$) is linearly dependent on the short-circuit current. We attribute this tunable $V_{oc}$ to a small tunneling resistance times a nearly constant photocurrent generation, which can be modeled as an ideal current source in this limited bias range (Supplementary Information). The photocurrent will vary over a broader bias range as discussed later. In Fig. 2e, we report the photocurrent spectra measured in the bilayer and trilayer regions. Two peaks corresponding to the A and B exciton absorption are observed. The electron and hole will be separated into different layers under the DEP and become dissociated as free carriers via the interfacial charge transfer\cite{gregg2003excitonic} (Supplementary Information). The resemblance between photocurrent spectra and $\text{MoS}_2$ absorption confirms the photocurrent origin.

Similar to the Bernal stacked bilayer graphene \cite{ju2015topological}, there are two possible stacking domains in a 3R bilayer $\text{MoS}_2$ (Fig. 3a): In the AB domain, the molybdenum atom in the top layer is above the sulfur atom in the bottom layer, while in the BA domain, the same amount of shift happens along the opposite direction, rendering an in-plane mirror image. The spontaneous polarization direction and photocurrent polarity are opposite between these two domains. Among the ten devices we fabricated, most devices show homogeneous photoresponse (more device mappings are in the Supplementary Information), except one sample (D2) shows both positive and negative photoresponse, which we attribute to the coexistence of AB and BA domains. The photocurrent mapping of D2 is shown in Fig. 3a, where the AB-BA domains show nearly symmetric responses. The monolayer region in D2 has a close-to-zero photoresponse, in agreement with D1. 

Made of atomically thin materials, the doping level of our device can be electrostatically tuned. Since the Dirac point of graphene is in the middle of $\text{MoS}_2$ bandgap, most carriers are doped into graphene rather than $\text{MoS}_2$ (Supplementary Information). The back-gate voltage dependence of the zero-bias photocurrent in D2 is shown in Fig. 3b. In both AB and BA domains, the photoresponse drops quickly with increasing positive voltage when the top and bottom graphene are electron doped. Due to the weak screening in graphene, the top electrode can be doped to have a smaller but same order-of-magnitude amount of charge compared with the bottom electrodes\cite{britnell2012field,britnell2013strong}. According to our electrostatic model, the Fermi level of the bottom (AB domain) and top (BA domain) graphene can be raised to 230 and 150 meV above the Dirac point respectively, when a $V_g=8$ V is applied (Supplementary Information). On the other hand, the decrease is much slower in the negative voltage range when the graphene is hole doped (In a small negative range, the AB domain has an increasing photocurrent with the increasing voltage, which we attribute to the decrease in the photo-thermal electric effect as discussed in detail in the Supplementary Information).The similar back-gate dependence in AB and BA domains suggests that the photoresponse change is not induced by an electric field in $\text{MoS}_2$\cite{yu2013highly,britnell2013strong} but by a doping effect in graphene - the extra charges doped into the graphene electrodes shift the chemical potential and occupy the states to which the photocarriers can be transferred to. This trend is reproducible in other single-domain devices we studied.

The asymmetric response of D2 to the electron and hole doping suggests our photocurrent is dominated by the electron transfer process. In bilayer $\text{MoS}_2$, the highest valence band edge is at the $\Gamma$ point while the K and Q points share a similar energy which is lowest in the conduction band. More importantly, the coupling between the top and bottom layer at the K and Q point is either strictly zero or comparable to the DEP induced interlayer potential, leading to a complete or significant layer polarization in the conduction band edge\cite{sung2020broken}. In each valley, the lower conduction band has more contribution from the layer with a lower potential, and the higher conduction band has a larger weight in the other. Initially, electrons are excited in both layers through inter-band transitions. After quickly relaxing to the lowest conduction band edge, they acquire a corresponding layer polarization, a process we call interlayer relaxation. Since the tunneling probability decreases exponentially with distance, the electrons in the top $\text{MoS}_2$ layer mostly tunnel to the top graphene, while those in the bottom layer mostly tunnel to the opposite electrode. Consequently, the layer polarization in the conduction band leads to a net imbalance between the two counter charge flows, and ultimately, to a PV current. An efficient PV effect based on this mechanism requires that the interlayer relaxation happens at a similar or faster rate than the charge tunneling to graphene. On the other hand, at the $\Gamma$ point, where most photo-excited holes relax to, the interlayer coupling is too large, about 400 meV, which is much larger than the DEP potential and leads to a negligible layer polarization\cite{sung2020broken}. Therefore, the holes contribute much less to the photocurrent. Together with the band filling effect in graphene, the asymmetric band structure in bilayer $\text{MoS}_2$ can explain the asymmetric doping dependence in the experiment.

Using this electron transfer picture, we can also understand the device's performance over a broad bias range. Since the tunneling current becomes significant under large bias, we define the photocurrent (PC) as the difference of current measured between laser on and off conditions. As shown in Fig. 3c and d, the PC is highly non-monotonically dependent on the bias. Across the full measurement range, the PC changes sign multiple times, showing negative slopes at the large bias ends. The response of the AB and BA domains are not symmetric either. We attribute this complicated bias dependence to the PV effect, plus the photo-thermal electric (PTE) ($I_{PTE}$) and bolometric (BOL) ($I_{BOL}$) effects in device\cite{xu2010photo, freitag2013photoconductivity,koppens2014photodetectors}. Here we leave the detailed discussion of thermal effects in the Supplementary Information and focus on the PV effect.

The impact of the external bias on the PV effect can be modeled by a two-band Hamiltonian (Supplementary Information). When the bias field is parallel to the DEP, the electrostatic potential difference between the top and bottom layer increases and the layer polarization is enhanced. If the bias is anti-parallel to the DEP, the layer polarization is reduced and can even be reversed. After fitting the experimental data using $I_{PC}=I_{PV}+I_{BOL}+I_{PTE}$, the pure PV contribution ($I_{PV}$) is extracted and shown in Fig. 3. (The thermal contributions of $I_{BOL}$ and $I_{PTE}$ are shown in Extended data Fig. 1.)

The PV effect accounts for most of the PC at zero bias and the saturation behavior at high bias. A compensation voltage, defined as the bias level when the PV current diminishes, is observed at 0.30($\pm$0.01) V and independent of the light intensity, a hallmark of the DEP-induced PV effect\cite{qin2008high}. The extracted compensation voltage and short-circuit current are symmetric between the AB and BA domains. The experimental compensation voltage is a few times larger than the spontaneous polarization induced interlayer potential from first-principles calculations\cite{park2019optical,sung2020broken}, potentially due to only a fraction of the applied bias drops across the tunneling junction in the experiment (The device tunneling behavior is confirmed by the gate dependence of I-V curves in Supplementary Information). The spontaneous layer polarization at zero bias, determined by the ratio between the experimentally measured interlayer potential and interlayer coupling, is 66($\pm$3)\% top and 34($\pm$3)\% bottom for the AB domain, which is close to the theoretical value predicted for the Q valley (59.1\% top and 40.9\% bottom)\cite{sung2020broken}. This agreement suggests the major contribution to the photocurrent arises from the photocarrier in the Q valley, likely due to its extra valley degeneracy compared to the K valley. As the bias voltage becomes much larger than the interlayer coupling, the band edge becomes completely layer polarized and the AB-BA domains show identical saturation behavior. Overall, our model captures the main features of the bias dependence and confirms a dominant PV effect arising from the photo-excited electrons.

In contrast with 3R, the mirror symmetry is restored in the 2H-$\text{MoS}_2$. Here we study a 2H bilayer sample with the same configuration as a control device (C1). In C1, we observe a negligible PC at zero bias and the EQE is about two orders of magnitude smaller than D2 (Extended data Fig. 2). The bias dependence is also totally different. The PC increases monotonically with external bias, which is known to be caused by the bias-induced tunneling barrier difference between the top and bottom electrodes\cite{yu2016unusually}. Without any spontaneous polarization, the top and bottom graphene are undoped with a symmetric chemical potential, and thus are free of the PTE and BOL effects.

As a commensurate crystal structure, the 3R stacking and its spontaneous polarization induced PV effect can be scaled up beyond bilayer. By assuming a constant polarization, we calculate the DEP in a graphene-contacted device with different $\text{MoS}_2$ thickness (Fig. 4a). In agreement with the conventional ferroelectric tunneling junction, the DEP decreases with the thickness\cite{mehta1973depolarization}, but the decrease is slower than 1/(L-1), since the graphene doping increases with thickness. As a result, the DEP-induced potential difference between the top and the bottom interfacial $\text{MoS}_2$ layers increases with thickness, leading to an increasing layer polarization. This enhanced layer polarization in combination with the higher optical absorption in the thicker device should improve the PV efficiency. In the thick limit, other factors such as the optical skin depth, Schottky junction effects, and imperfect stacking order need to be taken into account to estimate the ultimate PV efficiency. In in-plane devices, the DEP and its related PV effect become negligible as the polar material becomes a few $\mu$m long\cite{zhang2019enhanced,akamatsu2021van}.

To verify the layer dependence prediction, we fabricated another device with about ten layers of 3R-$\text{MoS}_2$ (D3). Compared with D2, the photoresponse of D3 is significantly stronger, reaching a responsivity of 70 $\text{mAW}^{-1}$ or an EQE of 16$\%$. The high efficiency suggests the interlayer relaxation within $\text{MoS}_2$ is no slower than the ultrafast interfacial charge transfer between $\text{MoS}_2$ and graphene\cite{jin2018ultrafast}. In contrast with previous bulk photovoltaic devices based on $\text{WSe}_2$\cite{akamatsu2021van}, our photocurrent increases linearly with the laser power up to $\sim$ mW level and does not become square-root power dependent in the saturation regime, confirming again a different origin. The estimated shift current response in 3R-$\text{MoS}_2$ bilayer along the out-of-plane direction is one to two orders smaller than what we observed in the experiments (D1, D2, and D4) (Supplementary Information), indicating the shift current is not the dominant mechanism in our device.

Finally, we benchmark our device against other unconventional photovoltaic devices. Compared with the most efficient devices based on the bulk photovoltaic effect\cite{zhang2019enhanced,akamatsu2021van}, ours is over ten times more efficient, polarization-independent, and gives a much larger photocurrent before saturation. Despite not being ideal for energy-harvesting applications due to its negligible $V_{oc}$, our device has a large zero-bias EQE, which is comparable with the other TMD-based photo-detectors relying on the extrinsic electric fields, either induced by the gate voltage\cite{yu2013highly,britnell2013strong} or in the p-n and Schottky junctions\cite{lee2014atomically,yu2013highly}. Compared with these extrinsic effects, our device benefits from requiring zero external voltage and involving less interfaces.The scalability, fast carrier extraction, and potentially switchable spontaneous polarization make it possible to build memory-integrated photodetectors based on 3R-$\text{MoS}_2$\cite{woods2021charge,yasuda2021stacking} 
Although similar DEP induced PV effect has been observed in polar oxides\cite{qin2008high,nechache2015bandgap}, our device explores the atomically thin limit of this effect. The exciton-enhanced light-matter interaction and smaller band gaps in TMDs also make them more appealing for many optoelectronic applications\cite{mak2016photonics, wang2018colloquium}.

\section*{M\lowercase{ethods}}
\textbf{Sample fabrication}:The 3R-$\text{MoS}_2$ bulk crystal is grown by chemical vapor transport method and characterized by XRD as discussed in our previous work\cite{suzuki2014valley}. All flakes are exfoliated on either Si/SiO2  substrates. Graphene electrodes are cut into specific shapes with an ultrafast laser. The flake thickness is identified by optical microscope and linear absorption spectroscopy\cite{zhao2016atomically}(Fig. S1). The 3R phase is confirmed by SHG measurement (Fig. S1). The thickness of the D3 flake is measured by AFM. Exfoliated hexagonal boron nitride (hBN) is used for encapsulation. All devices are fabricated using dry transfer method under the ambient condition\cite{wang2013one}. During the fabrication, flakes are aligned using a home-built transfer stage to avoid direct contact between top and bottom graphene. The electrical contact is achieved through overlapping the graphene with gold electrodes pre-patterned by optical lithography on heavily p-doped Si/$\text{SiO}_2$ substrates.

\textbf{Photocurrent measurement}: Our photocurrent is measured by a home-built scanning microscope with a 100x objective lens (0.75 NA). A 532nm laser is normally incident on the sample area at room temperature. The diffraction-limited laser focus spot is estimated to be 0.7 $\mu$m. The small range I-V curve in Fig. 2 is measured by a sourcemeter. Other photocurrent measurements are carried out through lock-in detection. The laser is chopped at about 900 Hz and the photocurrent is measured using a lock-in amplifier in series with a sourcemeter, which is used to compensate the lock-in input offset voltage and apply bias. The gate voltage is applied between the bottom graphite and bottom graphene electrode while the bias voltage is applied between the top and bottom graphene using two sourcemeters. The bias voltage scan range in Fig. 3c and 3d is from -1.5 V to +1.5 V, with a step size of 0.1 V. Each voltage scan takes 10s. We averaged for 10 times in the small laser power range (10-20 $\mu$W) to improve the signal to noise ratio and averaged for 2 times in the large power range (35-72 $\mu$W). The photocurrent bias dependence is repeatable between measurements with no hysteresis observed. The wavelength dependence of the photocurrent is measured with a monochromatized supercontinuum laser, from 1.6 to 2.4 eV, with a 100 ps pulse duration and 1 MHz repetition rate. Photoresponsivity is calculated from the photocurrent normalized to the average power, measured by a silicon power meter. The gate dependence of the photocurrent is measured with a He-Ne laser of 633 nm in order to minimize the photodoping effect\cite{ju2014photoinduced}.

\textbf{Acknowledgement:} Z.Y., D.Y., J.W.,J.L, B.Z, M.F., T.S. and K.A. acknowledge support from the 
Natural Sciences and Engineering Research Council of Canada,
Canada Foundation for Innovation, New Frontiers in Research
Fund, Canada First Research Excellence Fund, and Max
Planck–UBC–UTokyo Centre for Quantum Materials. Z.Y. is also
supported by the Canada Research Chairs Program. B.Z and M.F. acknowledge Quantum Materials and Future Technologies Program, and the Croucher Foundation.
Y.I. acknowledges support from JSPS Grant-in-Aid for
Scientific Research (S) (JP19H05602) and the A3 Foresight Program. T.I. acknowledges Grant-in-Aid for Scientific Research on Innovative
Areas (JP20H05264), Grant-in-Aid for Scientific Research (B)
(JP19H01819) and JST PRESTO (JPMJPR19L1). K.W. and T.T. acknowledge  support by the Elemental Strategy Initiative of MEXT, Japan and CREST (no. JPMJCR15F3), JST. The authors would like to thank Jerry Dadap, Zefang Wang, Joshua Folk, David Jones, George Sawatzky, Andrea Damascelli, and Ting Cao for helpful discussion.

\textbf{Author Contributions:} Z.Y. and D.Y.conceived this work. D.Y.,T.S. and K.A. fabricated the sample. D.Y., J.W. and J.L. conducted the measurement under supervision from Z.Y.. B.Z. performed the theory calculation under supervision from M.F. and Z.Y. . T.I., Y.I., K.W. and T.T. provided the bulk crystal. Z.Y., D.Y. and J.W. analysed the data. Z.Y. and D.Y. wrote the manuscript based on the input from all other authors. D.Y. and J.W. contributed equally to this work.

\textbf{Competing interests:} The authors declare no competing interests.

\textbf{Data and materials availability:} The data that support the plots
within this paper and other findings of this study are available from
the corresponding author upon reasonable request.

\textbf{References}
\makeatletter 
\renewcommand\@biblabel[1]{#1.} 
\makeatother %
\bibliographystyle{naturemag}

\bibliography{main}

\section*{F\lowercase{igures}}

\subsection*{Main data}
\subsection*{Fig.1}

\begin{figure}[H]
\centering
\includegraphics[width=0.6\textwidth]{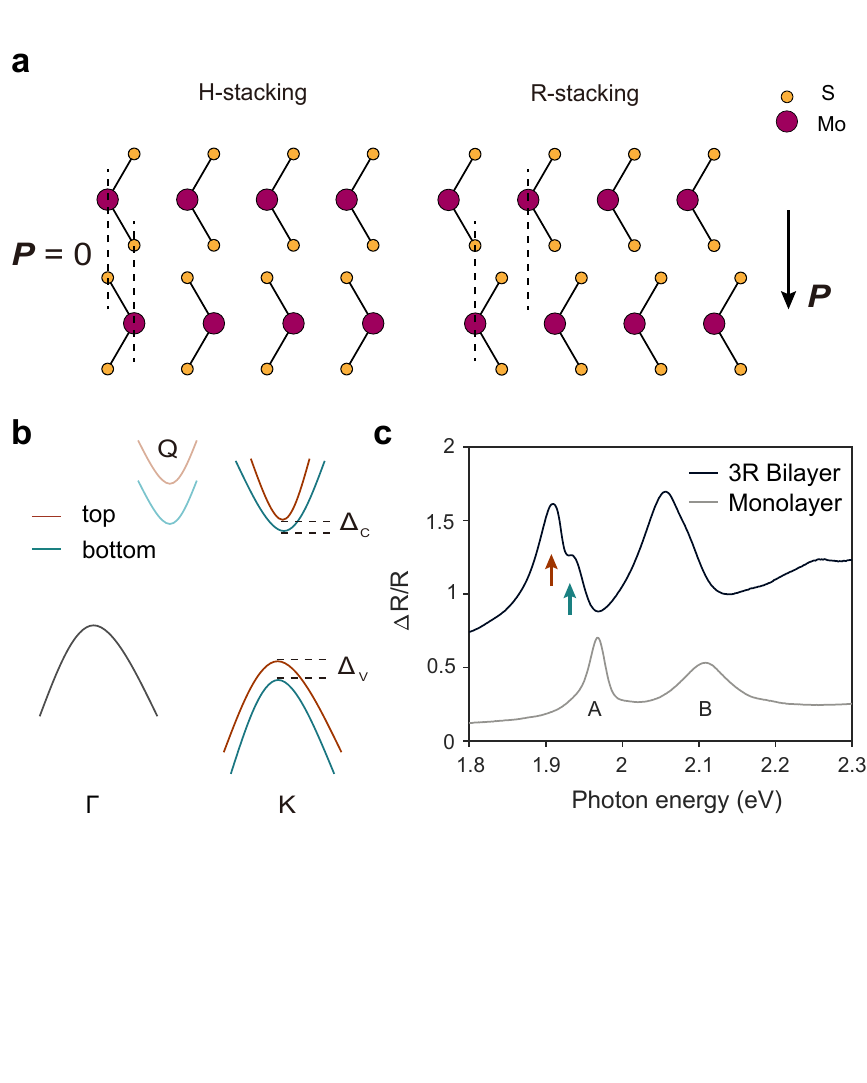}
\caption*{\textbf{FIG.1:} \textbf{a,} Schematics of H-stacking (2H) and R-stacking (3R) of bilayer $\text{MoS}_2$. The yellow and magenta colored balls correspond to the S and Mo atoms, respectively. The black arrow represents a spontaneous polarization between layers. \textbf{b,} Schematic of the band structure of a 3R-$\text{MoS}_2$ bilayer. Brown and green bands represent the top and bottom layers' contribution respectively. The smallest band gap is an indirect type with the highest valence band edge at the $\Gamma$ point. At the $K$ point,the conduction and valence bands in the two layers are decoupled. Finite splittings of $\Delta_{c}$ and $\Delta_{v}$ are a manifestation of the spontaneous polarization. The bands at $\Gamma$ and $Q$ points are layer coupled but only the $Q$ point is highly layer polarized as a result of the depolarization field. \textbf{c,} Reflection contrast spectra of monolayer (grey) and bilayer 3R-$\text{MoS}_2$ (black). The bilayer spectrum is shifted upward by 0.6 for clarity. The monolayer only shows two prominent peaks of A and B excitons. In contrast, in the bilayer case, the A exciton resonance splits into two peaks, corresponding to the inter-band transitions in the top (brown arrow) and bottom (green arrow) layer due to the asymmetric splitting in conduction and valence bands ($\Delta_{c} \neq \Delta_{v}$).}
\end{figure}

\subsection*{Fig.2}
\begin{figure}[H]
\centering
\includegraphics[width=1\textwidth]{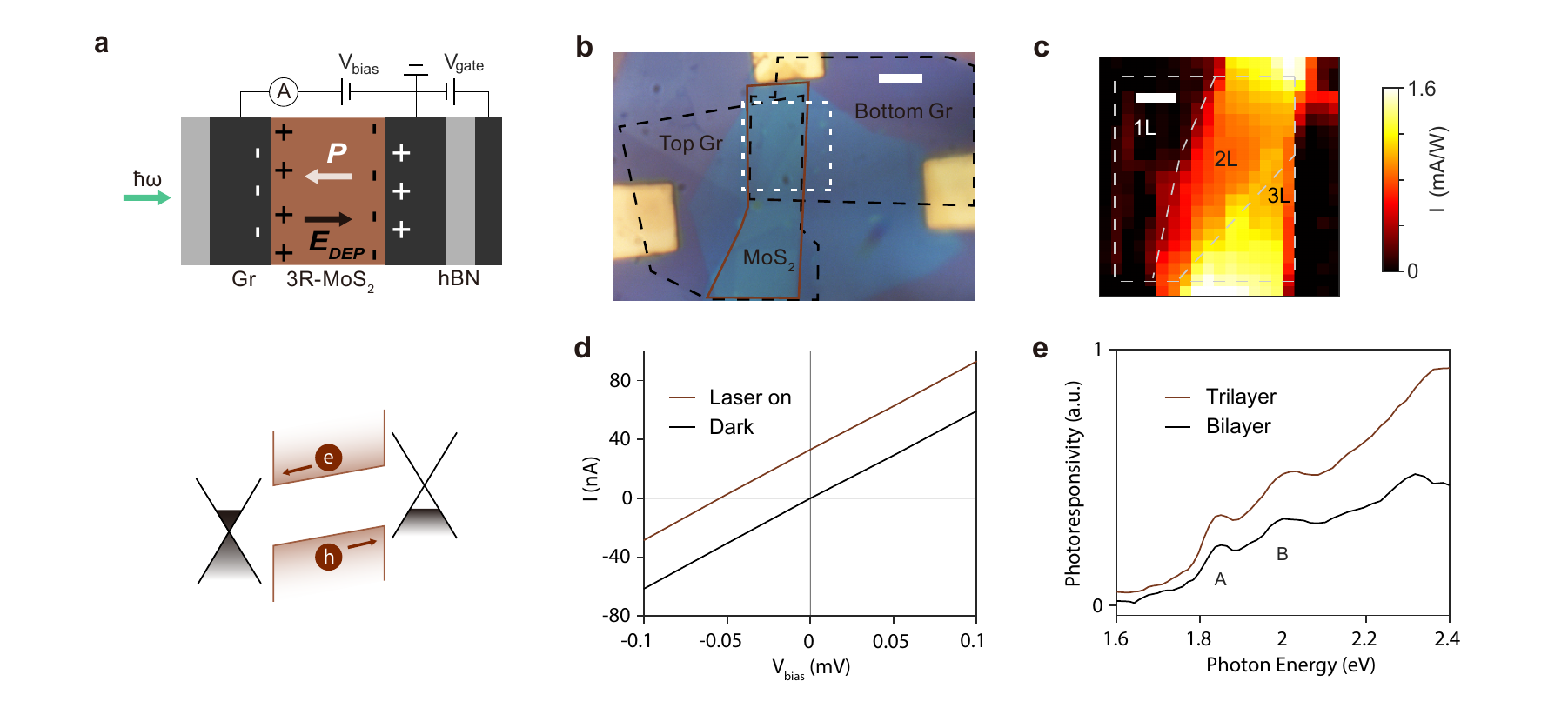}
\caption*{\textbf{FIG.2:} \textbf{a,} The top panel is a schematic of the tunneling junction in our device. The spontaneous polarization (white arrow) leads to polarization charges at the interface (black), which induces image charges in the graphene electrodes (white). Since the image charges do not fully compensate the polarization charges, a depolarization field emerges (black arrow). The bottom panel illustrates the band alignment in the device. The top and bottom graphene are electron and hole doped by the same amount of image charges. Photocarriers generated in the $\text{MoS}_2$ are driven to the interface and collected by graphene electrodes. \textbf{b,} Optical image of the D1 device, consisting of one, two, and three layers within the white dashed box. Graphene and 3R-$\text{MoS}_2$ are outlined by black dashed line and brown solid line. The scale bar is 5 $\mu m$. \textbf{c,} Scanning photocurrent map of D1, corresponding to the white dashed box of \textbf{b}. Monolayer, bilayer, and trilayer regions are indicated. The scale bar is 2 $\mu m$.  \textbf{d,} Current-voltage (I-V) curves of the bilayer region as measured in dark (black) and illuminated conditions. ($\lambda$ = 532 nm, brown). \textbf{e,} Photoresponsivity spectra of the bilayer and trilayer region. The two peaks near 1.8 and 2 eV correspond to the A and B excitons, respectively.}
\end{figure}

\subsection*{Fig.3}
\begin{figure}[H]
\centering
\includegraphics[width=1\textwidth]{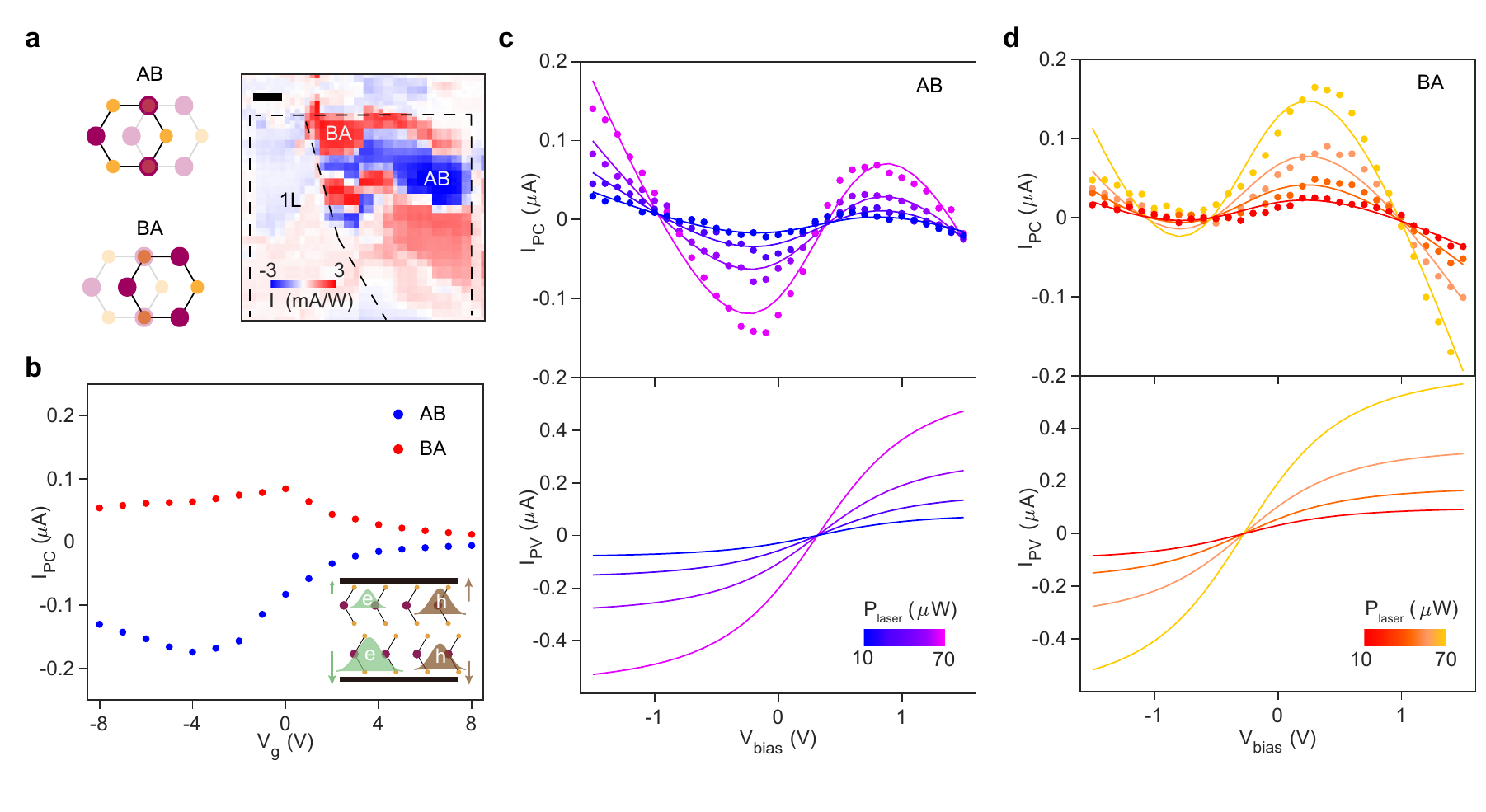}
\caption*{\textbf{FIG.3:} \textbf{a,} Left panel is a schematic of two possible stacking domains (AB and BA) of a 3R bilayer $\text{MoS}_2$. Top layer (solid) shifts towards left (AB) and right (BA) respectively, relative to bottom layer (translucent). Right panel is the scanning photocurrent map of device D2. Positive and negative photoresponse areas correspond to AB and BA domains with almost symmetric responsivity ($\pm$ 3  $\text{mA/W}$). A monolayer region is adjacent to the bilayer domains, exhibiting negligible photoresponse. The scale bar is 1 $\mu$m. \textbf{b,} Back-gate voltage dependence of the photocurrent in the AB (Blue dots) and BA (red dots) domains. The inset of \textbf{b,} shows a schematic of charge transfer between 3R-$\text{MoS}_2$ and graphene (black). Electrons (green) are partially layer polarized while holes (brown) are equally distributed between two layers.  \textbf{c} and \textbf{d,} top panels are the bias dependence of the photocurrent (PC) in the AB (blue series) and BA (red series) domains. The dots represent the measured PC at each bias voltage at different laser powers between 10 and 70 $\mu$W. The solid lines are fits to the data based on the model of  $I_{PC}=I_{PV}+I_{BOL}+I_{PTE}$. $I_{PV}$ is the intrinsic photovoltaic effect. $I_{BOL}$ and $I_{PTE}$ are bolometric and photo-thermal electric effect, both of which are thermal contributions from graphene electrodes. Bottom panels are the extracted photovoltaic I-V dependence. The compensation voltage where photocurrent stops is $\pm $0.3 V in AB and BA domains.}
\end{figure}

\subsection*{Fig.4}
\begin{figure}[H]
\centering
\includegraphics[width=0.6\textwidth]{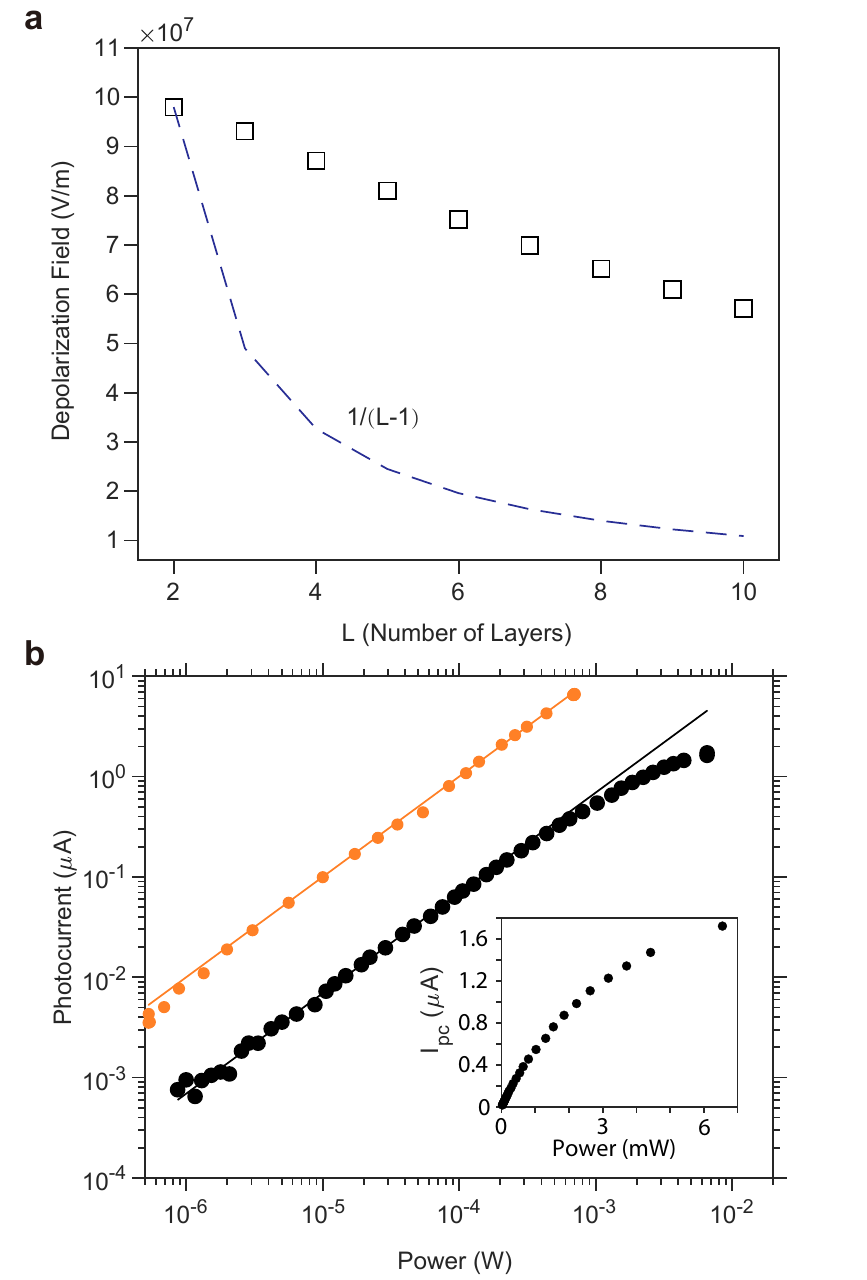}
\caption*{\textbf{FIG.4:} \textbf{a,} Simulated depolarization field (DEP) strength versus the $\text{MoS}_2$ thickness (L) assuming a thickness-independent polarization (black rectangular). The blue dash line decreases as 1/(L-1), which is faster than the calculated DEP-layer dependence, indicating a stronger response in the thicker device. \textbf{b,} Laser power dependence of the photocurrent at zero bias for device D2 (black) and D3 (orange). The dots are experimental data and the solid lines correspond to a linear dependence fit. The external quantum efficiency of D3 is one order of magnitude larger than that of D2. The inset shows the power dependence of D2 in the linear scale.}
\end{figure}

\subsection*{Extended data}

\subsection*{Extended data Fig.1}
\begin{figure}[H]
\centering
\includegraphics[width=0.8\textwidth]{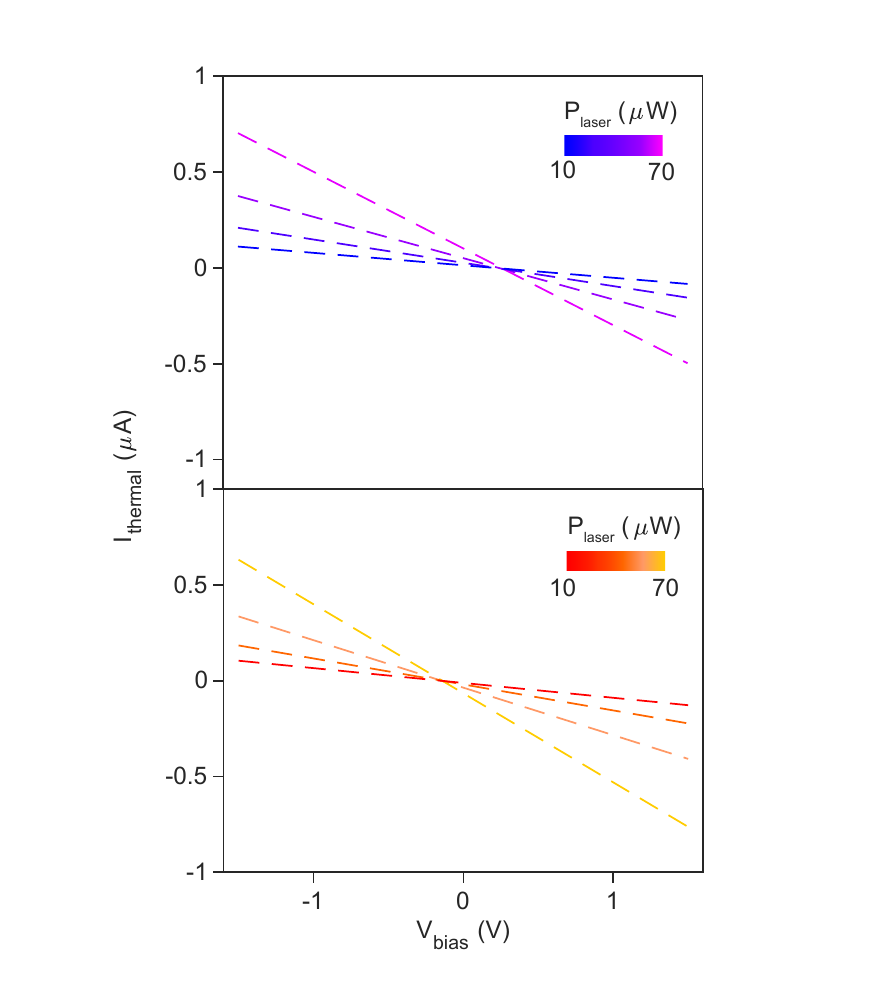}
\caption*{\textbf{Extended data Fig.1:} Extracted thermal contributions of the AB (top panel) and BA (bottom panel) domains in device D2. The straight lines with negative slopes are due to bolometric effect $I_{BOL}$, which crosses zero at zero bias. The non-zero intercept is caused by the photo-thermal electric effect $I_{PTE}$, which is always opposite in direction to $I_{PV}$. The $I_{PTE}$ is approximately independent of the bias.}
\end{figure}

\subsection*{Extended data Fig.2}
\begin{figure}[H]
\centering
\includegraphics[width=0.8\textwidth]{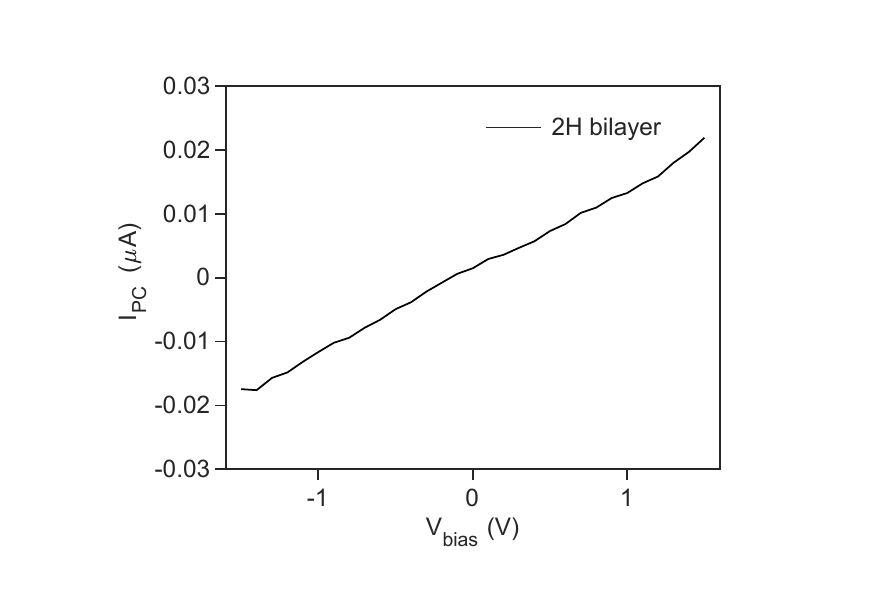}
\caption*{\textbf{Extended data Fig.2:} Bias dependence of the photocurrent in the 2H bilayer device, C1. With a similar laser illumination condition (P = 20 $\mu W$), C1 shows a near zero photocurrent under zero bias. The photocurrent linearly increases with the bias, with no thermal contribution observed.}
\end{figure}

\end{document}